\documentclass[pre,twocolumn,floatfix,superscriptaddress,longbibliography,nofootinbib]{revtex4-1}
\RequirePackage[sort&compress]{natbib}

\usepackage{bbold}
\usepackage{bbm}
\usepackage[pdftex]{graphicx}
\usepackage{latexsym,amsmath,verbatim}
\usepackage{color}
\usepackage{rotating}
\usepackage{multirow}
\usepackage[english]{babel}
\usepackage{color}
\usepackage{changes}
\usepackage{hyperref}
\usepackage{booktabs}

\usepackage{lineno}

\usepackage{subfigure}

\usepackage{latexsym,amsmath,verbatim}
\usepackage{mathrsfs}
\usepackage{mathtools}

\usepackage{boxedminipage}

\usepackage{multirow}

\usepackage{ifpdf}

\usepackage{bm}

\ifpdf
\usepackage[pdftex]{graphicx}
\else
\usepackage{graphicx}
\fi

\ifpdf 
\DeclareGraphicsExtensions{.pdf,.jpg,.png}
\else
\DeclareGraphicsExtensions{.eps,.jpg}
\fi

\begin{document}

\title{Quantifying the Diaspora of Knowledge in the Last Century}

\author{Manlio De Domenico$^{1}$, Elisa Omodei$^{2}$, Alex Arenas$^{1}$\\
\normalsize{$^{1}$Departament d'Enginyeria Inform\`{a}tica i Matem\`{a}tiques, Universitat Rovira i Virgili, 43007 Tarragona, Spain}}

\begin{abstract} 
Academic research is driven by several factors causing different disciplines to act as ``sources'' or ``sinks'' of knowledge. However, how the flow of authors' research interests -- a proxy of human knowledge -- evolved across time is still poorly understood. Here, we build a comprehensive map of such flows across one century, revealing fundamental periods in the raise of interest in areas of human knowledge. We identify and quantify the most attractive topics over time, when a relatively significant number of researchers moved from their original area to another one, causing what we call a ``diaspora of the knowledge'' towards sinks of scientific interest, and we relate these points to crucial historical and political events. Noticeably, only a few areas -- like Medicine, Physics or Chemistry -- mainly act as sources of the diaspora, whereas areas like Material Science, Chemical Engineering, Neuroscience, Immunology and Microbiology or Environmental Science behave like sinks.
\end{abstract}

\maketitle

\section{Introduction}

Nowadays, the research carried out by academics in all areas of human knowledge is heavily driven by exogenous factors, such as allocation of funding resources or political interests~\cite{boyack2003indicator,ma2015anatomy}. Two decades ago, pioneering studies by Etzkowitz and Leydesdorff already put in evidence the importance of relationships between university, industry and government~\cite{etzkowitz1995triple,leydesdorff1998triple,etzkowitz2000dynamics}, a ``triple helix'' that shapes and drives the development of knowledge, impelling researchers to change research interests or their institution~\cite{boyack2005mapping,leydesdorff2009global,deville2014career}. The structure and evolution of human knowledge has been extensively investigated by observing, for instance, how academics tend to choose their co-authors or they physically move between different research institutions~\cite{etzkowitz1995triple,leydesdorff1998triple,etzkowitz2000dynamics,shiffrin2004mapping,borner2004simultaneous,boyack2005mapping,leydesdorff2009global,deville2014career,ke2015defining,sinatra2015century}. These analyses, often based on citation patterns among authors, institutions, papers or journals, allow to understand how disciplines are related to each other in terms of scientific production and impact, but are not intended to quantify the flow of knowledge in science or to identifying crucial periods for the development of human knowledge. In fact, the interest of researchers are not rarely driven by currently available funding opportunities or by political choices, an emblematic example being the investments in nuclear physics during the World War II. Such factors, often external to the context of academy research, act as catalysts pushing researchers to leave their current area of interest towards different areas.

To study this phenomenon of ``knowledge diaspora'', we consider the Microsoft Academic Graph, a data set of more than 35,000,000 of papers published in more than 21,000 different journals in the last 100 years. We are interested in exploiting metadata information to classify each paper into one or more disciplines. Unfortunately, our exploratory analysis of the classification scheme released with the dataset, based on paper keywords, revealed some relevant drawbacks that would dramatically bias the more sophisticated analysis presented in this work. To cope with such limitations, we classified the papers according to the journal where they have been published, the rationale behind this choice being that journals tend to publish research studies that are, in general, more pertinent to their specific topic(s). For instance, it is difficult to publish in a physics journal a paper about humanities or biology, if this paper does not provide some physical insights that would make it suitable for an audience of physicists. Therefore, each journal is classified into one or more topics, fine-grained representations of academic knowledge, and into one or more areas, coarse-grained representations of academic knowledge. We use the SCImago classification, where there are 306 unique topics grouped into 27 distinct areas of knowledge to assign topics and areas to each paper, according to its journal. One possible cause of criticism might be that such classification is too recent to characterize adequately journals existing at the beginning of the past century. However, it must be remarked that we are focusing our attention on those journals that have a long-established tradition -- i.e., from a few decades up to one century -- and are unlikely to have dramatically changed their area of reference across years. We have divided the data set into non-overlapping temporal snapshots of 5~years, from 1910 to 2014. A snapshot marked with a year refers to a period between that year and 4 years later, e.g. 2000 refers to the period 2000-2004. To trace the changes in research interests of every author in the data set from one temporal snapshot to the successive, we count how many authors published in topic A at time $\tau$ and in the same or a different topic B at time $\tau+\Delta \tau$ (see Appendix). The volume of authors linking topics defines an evolving network of connections among topics, i.e. a multilayer (time-varying, weighted and directed) network~\cite{holme2012temporal,kivela2013multilayer,dedomenico2013mathematical,boccaletti2014structure}. The same procedure has been also applied to the coarser level of areas.
The structure of these dynamical multilayer networks encode the publishing temporal dynamics of academics who change their research interests across knowledge topics and areas, respectively. In the following we will simply refer to these structures using the term network, avoiding to specify that they are time-varying and multilayer.

\section{Overview of the data set} 

The Microsoft Academic Graph is a heterogeneous graph containing scientific publication records, citation relationships between those publications, as well as authors, institutions, journals and conference ``venues'' and fields of study~\cite{sinha2015overview}. We used the latest publicly available updated version (31 August 2015) of this data set\footnote{\url{http://research.microsoft.com/en-us/projects/mag/}} in our study. However, our careful inspection of the data did not allow us to use the accompanying classification of papers into fields of study. The first obstacle was the number of different keywords classifying the papers: tens of thousands of categories providing a scheme too fine-grained for our study. A reduction of such keywords into more general topics would require machine learning and heuristics that would introduce other uncontrollable bias in the resulting classification. The second obstacle was the unclear mechanisms adopted to assign one or more keywords to each paper. In fact, we have found many misclassified papers, an emblematic case being a paper about Agricultural Science that has been classified in several topics, among which General Relativity. Instead, we gathered data from an external (publicly available) source. More specifically, we used SCImago Journal and Country Rank in 2014\footnote{\url{http://www.scimagojr.com/journalrank.php}} to classify journals into 306 distinct research topics and 27 unique knowledge areas. Successively, we filtered out from the Microsoft Academic Graph data set all the papers that were not published in journals, thus excluding other venues such as conferences, and in particular we filtered out those papers published in journals that were not found in the SCImago classification. More than 35 millions of papers survived this filtering procedure, representing a promising 28.7\% of the original data set, and more than 60\% of the original set of papers published in journals only. The number of different journals matching the SCImago data set was 21,729, and we report in Supplementary Table~\ref{tab:journal-mux} some information about the distribution of their multiplexity, i.e. the number of different topics and areas where they are classified. If a journal is classified into just one topic or area, the papers published in that journal will not be classified as multiplex, whereas, conversely, papers published in journals that are classified into more than one topic or area, will be treated as multiplex. Finally, it is worth remarking that we further reduced the dataset to avoid the effects of non-disambiguated authors. More specifically, we built the distribution of the number of papers per year of each author and we focused on the 99.9\%-quantile distribution, i.e. we excluded the 0.1\% of authors. This choice excluded all the names who authored more than 17 journal papers per year, the rational being that names with a higher number of papers per year probably corresponds to different authors having the same name.

\begin{table}
\centering
\begin{tabular}{|cc|}
\hline
\textbf{\# of Topics} & \textbf{\% of Journals}\\
\hline
1 & 31.6\%\\
2 & 33.4\%\\
3 & 19.8\%\\
4 & 9.2\%\\
5 & 3.3\%\\
\hline
\end{tabular}
\begin{tabular}{|cc|}
\hline
\textbf{\# of Areas} & \textbf{\% of Journals}\\
\hline
1 & 50.9\%\\
2 & 36.2\%\\
3 & 9.8\%\\
4 & 2.1\%\\
5 & 0.5\%\\
\hline
\end{tabular}
\caption{\label{tab:journal-mux}Multiplexity of journals with respect to topics and areas. We report the percentage of journals that are classified by SCImago in exactly 1, 2, ..., \emph{etc}, topics (areas).
Only statistics for the top five are reported, with rapidly decreasing percentage of journals classified in more than five topics (areas).}
\end{table}

\section{Multilayer network model} 

The data set used in our study contains a huge amount of information about published papers and their authors. We focused on specific subsets of the data, including author name, the papers he/she published, the journal where they have been published and the publishing year. Thanks to the SCImago classification of areas of knowledge, we were able to assign one or more topics to each journal. Thus, we built a tripartite time-varying multilayer network $\mathcal{G}$ where for each temporal snapshot $\tau$, a tripartite multiplex $\mathcal{M}$ is considered. Each multiplex is composed by layers $\mathcal{L}$ -- identifying topics or areas of knowledge, depending on the application of interest -- where there are three types of nodes: authors (A), papers (P) and journals (J). One or more authors are linked to the paper(s) they co-authored that, in turns, are linked to the journal where they have been published, resulting in a bipartite network linking nodes of type A to nodes of type P, and a bipartite network linking, at the same time, nodes of type P to nodes of type J. If a journal is classified in more than one topic or area, the links are replicated accordingly across layers. The resulting network is tripartite, because three types of nodes are involved, and multiplex, because nodes are replicated on different layers. For our purposes, we aggregated the tripartite network in each layer $l\in\mathcal{L}$ with respect to papers, in order to obtain multiplex bipartite networks of authors and journals only, for each temporal snapshot. Finally, each node is inter-connected to its replicas in other layers and temporal snapshots. The mathematical representation~\cite{dedomenico2013mathematical,kivela2013multilayer} of $\mathcal{G}$ is a rank-6 tensor $G^{\alpha\tilde{\gamma}\bar{\epsilon}}_{\beta\tilde{\delta}\bar{\phi}}$, where indices $(\bar{\epsilon},\bar{\phi})$ identify the temporal snapshot, $(\tilde{\gamma},\tilde{\delta})$ identify the layers and $(\alpha,\beta)$ identify the nodes.

This complex network, however, is not the final object we worked with. In fact, our analysis is more focused on changes in publication patterns across years. Mathematically, this means that we are more interested in the links between authors and journals exhibit between one temporal snapshot and the successive, i.e. in inter-layer links with respect to time. We derived a more suitable time-varying multilayer network $\mathcal{H}$ from $\mathcal{G}$ as follows. Let $A_{i}$ be the $i-$th node of type A (i.e. authors) and $J_{k}$ be the $k-$th node of type J (i.e. journals), regardless of topics (areas) classification and time. In $\mathcal{G}$, a link between $A_{i}$ and $J_{k}$ in layer $l$ at time $\tau$ exists if $G^{il\tau}_{kl\tau}>0$. Similarly, if in the successive snapshot $\tau'>\tau$ the same author $A_{i}$ is linked to journal $J_{k'}$ ($k'$ can be the same as $k$) in layer $l'$ ($l'$ can be the same as $l$), then $G^{il'\tau'}_{k'l'\tau'}>0$. Clearly, an author might publish papers on different topics or areas at time $\tau$ but he/she will be, in general, more active on one or a few more. For this reason for each snapshot, we will consider only the layer where the author has been more active, i.e. where $G^{il\tau}_{kl\tau}$ is maximum with respect to $l$ (note that if there is more than one layer where the author is equally active, we will consider all of those layers). We will indicate by $l^{\star}$ such layers. The components of the tensor representing $\mathcal{H}$ that encode inter-snapshot connections, are defined by
\begin{eqnarray}
H^{il^{\star}\tau}_{il'^{\star}\tau'} = \Theta(G^{il^{\star}\tau}_{kl'^{\star}\tau}) \times \Theta(G^{il'^{\star}\tau'}_{k'l'^{\star}\tau'}),
\end{eqnarray}
i.e. an interconnection between an author at time $\tau$ and his/her replica at time $\tau'>\tau$ is present if and only if the author published at time $\tau$ and at time $\tau'$. It is worth remarking that the replicas being linked are defined on layers $l^{\star}$ at time $\tau$ and $l'^{\star}$ at time $\tau'$, thus also connecting (possibly different) topics or areas across time. The presence of Heaviside step function $\Theta(\cdot)$ is to guarantee that each author is counted just once at this step, regardless if he/she produced more papers. It is evident that information about the flow of authors moving from one knowledge topic (or area) to another across time is only encoded in inter-snapshot connections among author's replicas, whereas the presence of journals as nodes is no more required, as well as intra-snapshot links, i.e. connections within the same temporal snapshot. Therefore, the tensor $H$ representing $\mathcal{H}$ is defined on a smaller tensorial space with respect to $G$, because nodes are just authors instead of authors and journals. Moreover, it is also extremely sparse and, in fact, it can be further aggregated without loss of information, because of the absence of intra-snapshot links, by projecting the tensor into the space of topics (or areas) and time, getting rid of information about authors (see Appendix for details about this step). The resulting tensor $M^{\tilde{\gamma}\bar{\epsilon}}_{\tilde{\delta}\bar{\phi}}$, that is the one we used in our analysis, represents a multilayer network where nodes are topics (or areas), identified by indices $(\tilde{\gamma},\tilde{\delta})$, and layers are temporal snapshots, identified by indices $(\bar{\epsilon}, \bar{\phi})$. Intra-layer links, i.e. connections among topics within the same temporal snapshot, are not present, whereas inter-layer links among topics encode the underlying flow of authors during consecutive periods of time.

\section{Results}

\begin{figure*}[!ht]
\centering
\includegraphics[width=16cm]{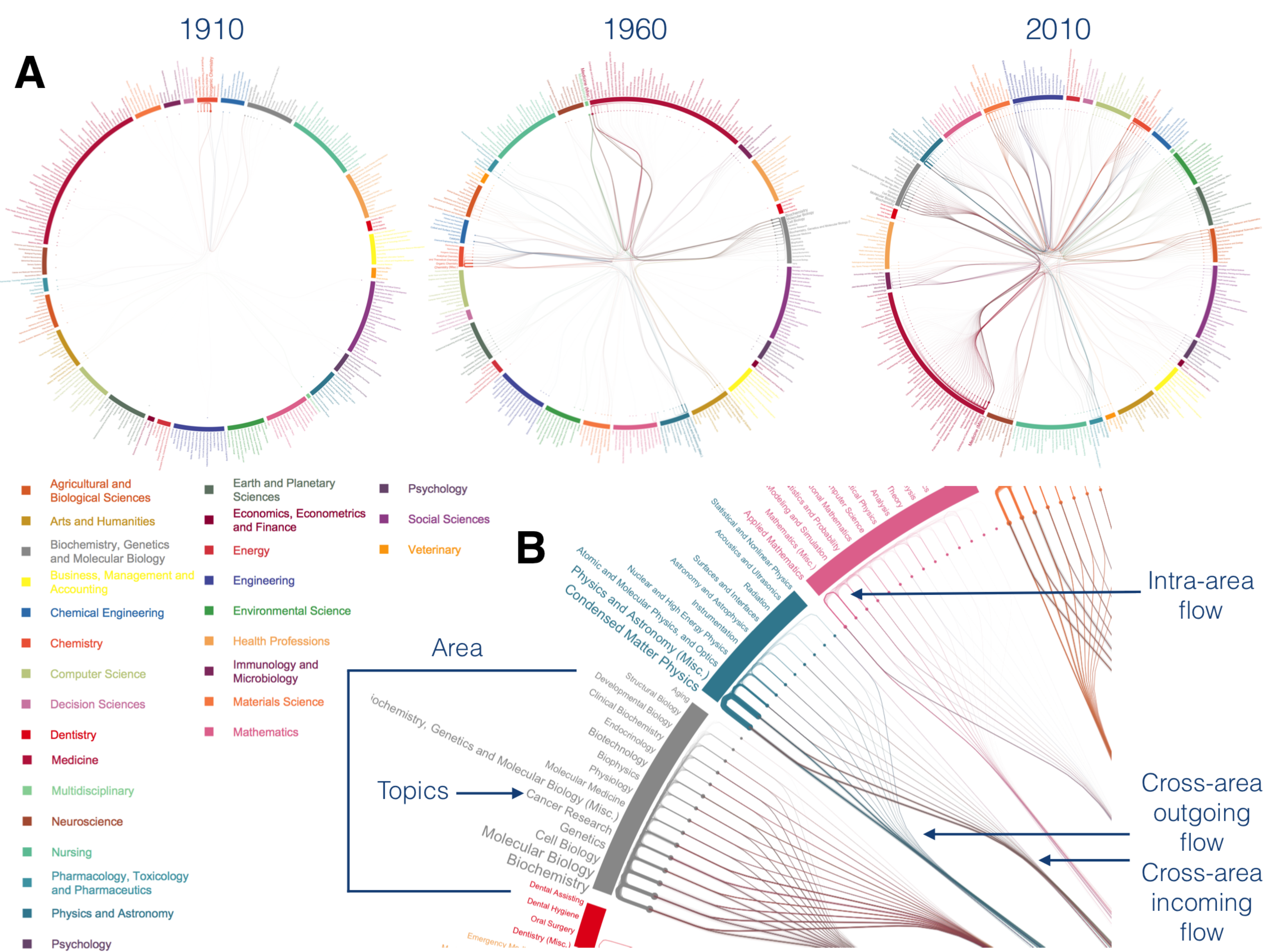}
\caption{\textbf{Flow network of knowledge diaspora.} Points on the circle indicate topics (fine-grained knowledge representations) that are colored according to their SCImago area (coarse-grained knowledge representations), represented by thick sectors, whose color legend is reported. Two topics are connected if at least one author at time $\tau$ switched from one to another 5~years later. (A) Flow of authors moving his/her research activity from one topic to others across time. (B) How to read this visualization: switches between topics of the same area, namely ``intra-area flows'', are represented as `U'' shaped links close to sectors, to distinguish them from ``cross-area flows''. The outgoing flow is colored by the area of origin. The width of edges is proportional to the observed flow. See Appendix for more details about topics classification and this type of visualization.}\label{fig:Fig1}
\end{figure*}

To gain the first insights about the knowledge diaspora across topics, we developed an \emph{ad hoc} visualization (see Appendix) to put in evidence, for each topic, the intricate web of flows of authors incoming from and outgoing to other topics. 
We see in Fig.~1 a few emblematic cases corresponding to the diaspora observed in 1910, 1960 and in 2010, covering one century of academic publishing in all areas of knowledge. It is evident that one century ago authors were not contributing significantly outside their own area of expertise.
After 50 years the diaspora is more prominent, with intense flows between topics of different areas, such as --Medicine-- and --Biochemistry, Genetics and Molecular Biology--, between --Physics and Astronomy-- and --Earth and Planetary Science--, or between --Chemistry-- and --Chemical Engineering--.  After 100 years, the diaspora is extremely evident, affecting basically all areas of knowledge.

The map of knowledge diaspora shown in Fig.~1 allows to get qualitative insight about this phenomenon, although it does not allow to quantify, for instance, the raise of research interest in specific topics. 
We will focus first our study on the emergence of topics of interest, by analyzing the variation of their incoming flows. 
To this aim, we quantify the attractiveness of a topic $t$ through time $\delta_{t}(\tau)$, by tracking the evolution of the relative changes in the volume of authors $V_{tt'}(\tau)$ incoming from all other topics $t'\neq t$, at each temporal snapshot $\tau$:
\begin{eqnarray}
\delta_{t}(\tau)&=&\frac{1}{N_{t}-1}\sum\limits_{t'\neq t}\frac{V_{tt'}(\tau)-V_{tt'}(\tau-5~\text{years})}{V_{tt'}(\tau-5~\text{years})},
\label{delta_vol}
\end{eqnarray}
being $N_{t}=306$ the total number of topics considered. For each topic, it  quantifies the average net relative change in the incoming flow. 
This parameter is sensitive to changes in the flow from one topic to another, even when this flow is rather small compared to the total incoming flow.
Indeed, it might happen that a topic attracts a small flow of authors from many other topics or a huge flow of authors from a rather small set of other topics. The parameter $\delta_{t}(\tau)$ would detect both patterns and assign a similar score in the two cases. Other aggregated parameters, such as the relative change in the overall incoming flow per topic, are not able to capture this type of patterns, that would be inevitably hidden by larger flows with possibly less significant relative variations over time.

\begin{figure*}[!ht]
\centering
\includegraphics[width=16cm]{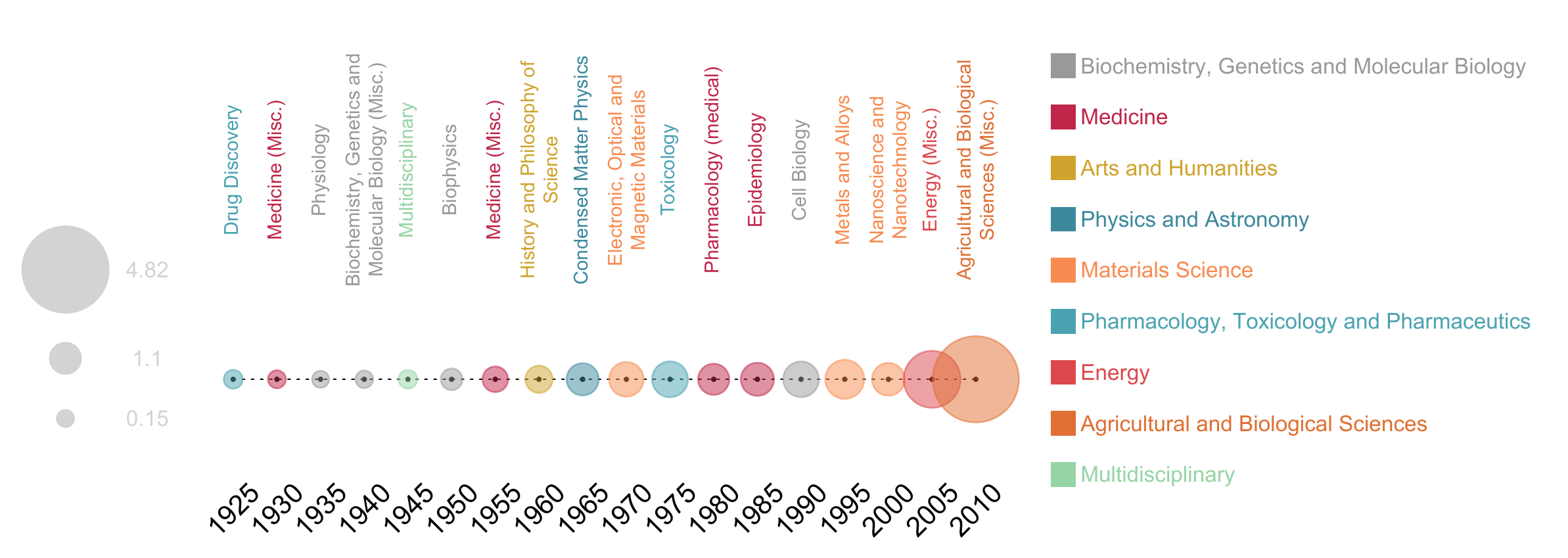}
\caption{\textbf{Most attractive topics in the knowledge diaspora.} The flow network of each temporal snapshot of 5~years is compared with the one immediately subsequent, and the relative changes in the volume of authors attracted by a topic (see Eq.\,(\ref{delta_vol})) are computed. For each temporal snapshot, we report the largest relative change observed in the volume. The relative increase is encoded in the size of circles.}\label{fig:Fig2}
\end{figure*}

For each snapshot $\tau$ separately, we look for the most attractive topic, the one with the highest value of $\delta_{t}(\tau)$. The results, shown in Fig.~2, reveal intriguing correspondences with historical or political events. For instance, between '60s and '70s the study of physical properties of liquids was officially included in solid state physics, to form the basis of Condensed Matter, name adopted in that period to redirected into one common field those physicists who were previously working on simple and complex matter~\cite{martin2015s}.

Another interesting case is represented by Nanotechnology, with a significant activity change between 2000 and 2004, following the Nobel Prize in Chemistry won by Harry Kroto, Richard Smalley, and Robert Curl for the discovery of fullerenes. Fundamentals in many technological applications, fullerenes attracted a large number of researchers from --Statistics and Probability--, --Modeling and Simulation-- and --Computer Science Applications--, when the new National Nanotechnology Initiative~(\url{http://www.nano.gov/}) was officially proposed (1999) and the US President Bill Clinton declared a budget worth \$500 million to support it (January 2000), thus justifying the diaspora from many other disciplines to --Nanoscience and Nanotechnology--. The case of --Agricultural and Biological Sciences (Misc.)--, exhibiting the largest value of $\delta_{t}(\tau)$ between 2010 and 2014, especially attracted our attention. A deeper analysis, revealed the presence of an increasing significant flow of researchers incoming from --Energy (Misc.)-- who moved their publications towards in journals pertaining to agricultural and biological sciences, with research about genetically modified organisms, synthesis of biomolecules, biofuels, food systems and bioenergy.

After the fine-grained analysis at the level of topics, we focus on the analysis at the coarse-grained level of areas. For the analysis at the area level we need to define the intra-area flow as the volume of authors $V^{\text{[intra]}}_{a}(\tau)$ that keep publishing in the same area $a$ over successive temporal snapshots. The overall cross-area incoming flow $V^{\text{[to]}}_{a}(\tau)$ is defined as the volume of authors who publish in area $a$ at time $\tau$ coming from other areas. Finally, the overall cross-area outgoing flow $V^{\text{[from]}}_{a}(\tau)$ is defined as the volume of authors in area $a$ that publish in other areas at time $\tau$. These measures allow to investigate many aspects of the diaspora, characterizing the role played by different areas in the evolution of human knowledge. We introduce two local descriptors, namely the immigration and the emigration indices defined by
\begin{eqnarray}
\iota_{a}(\tau)&=&\frac{V^{\text{[to]}}_{a}(\tau)}{V^{\text{[intra]}}_{a}(\tau)+V^{\text{[to]}}_{a}(\tau)}\\
\epsilon_{a}(\tau)&=&\frac{V^{\text{[from]}}_{a}(\tau)}{V^{\text{[intra]}}_{a}(\tau)+V^{\text{[from]}}_{a}(\tau)},
\end{eqnarray}
respectively, characterizing the diaspora from a local perspective, i.e. in terms of relative variations with respect only to the existing population of authors working in the area $a$. These indices range from 0 -- characterizing areas where the incoming (outgoing) flow of immigrating (emigrating) authors is negligible with respect to the existing authors population in the area -- to 1 -- indicating areas where the existing authors population is negligible with respect to the incoming (outgoing) flow of immigrating (emigrating) authors. However, these two local indices alone, do not allow to gain global insight about the diaspora from sources and to sinks of knowledge. For instance, such indices do not allow to understand if areas like --Physics and Astronomy--, --Mathematics-- or --Computer Science--, producing academics whose modeling and abstraction skills make them suitable for challenging problems in other disciplines, act as global sources of the diaspora or not. In fact, it might happen that even if academics from these areas are commonly perceived to be very multidisciplinary, their flow with respect to the intra-area flow of authors could be rather small. To this aim we introduce two global descriptors, namely the sink and source indices defined by
\begin{eqnarray}
\rho_{a}(\tau)&=&\frac{V^{\text{[to]}}_{a}(\tau)}{\sum\limits_{a'}V^{\text{[to]}}_{a'}(\tau)}\\
\sigma_{a}(\tau)&=&\frac{V^{\text{[from]}}_{a}(\tau)}{\sum\limits_{a'}V^{\text{[from]}}_{a'}(\tau)},
\end{eqnarray}
respectively. As before, such indices range from 0 -- indicating areas where the incoming (outgoing) flow of authors is negligible with respect to the overall incoming (outgoing) flow --  to 1 -- characterizing areas where the incoming (outgoing) flow of authors dominates the overall incoming (outgoing) flow.

\begin{figure*}[!ht]
\centering
\includegraphics[width=16cm]{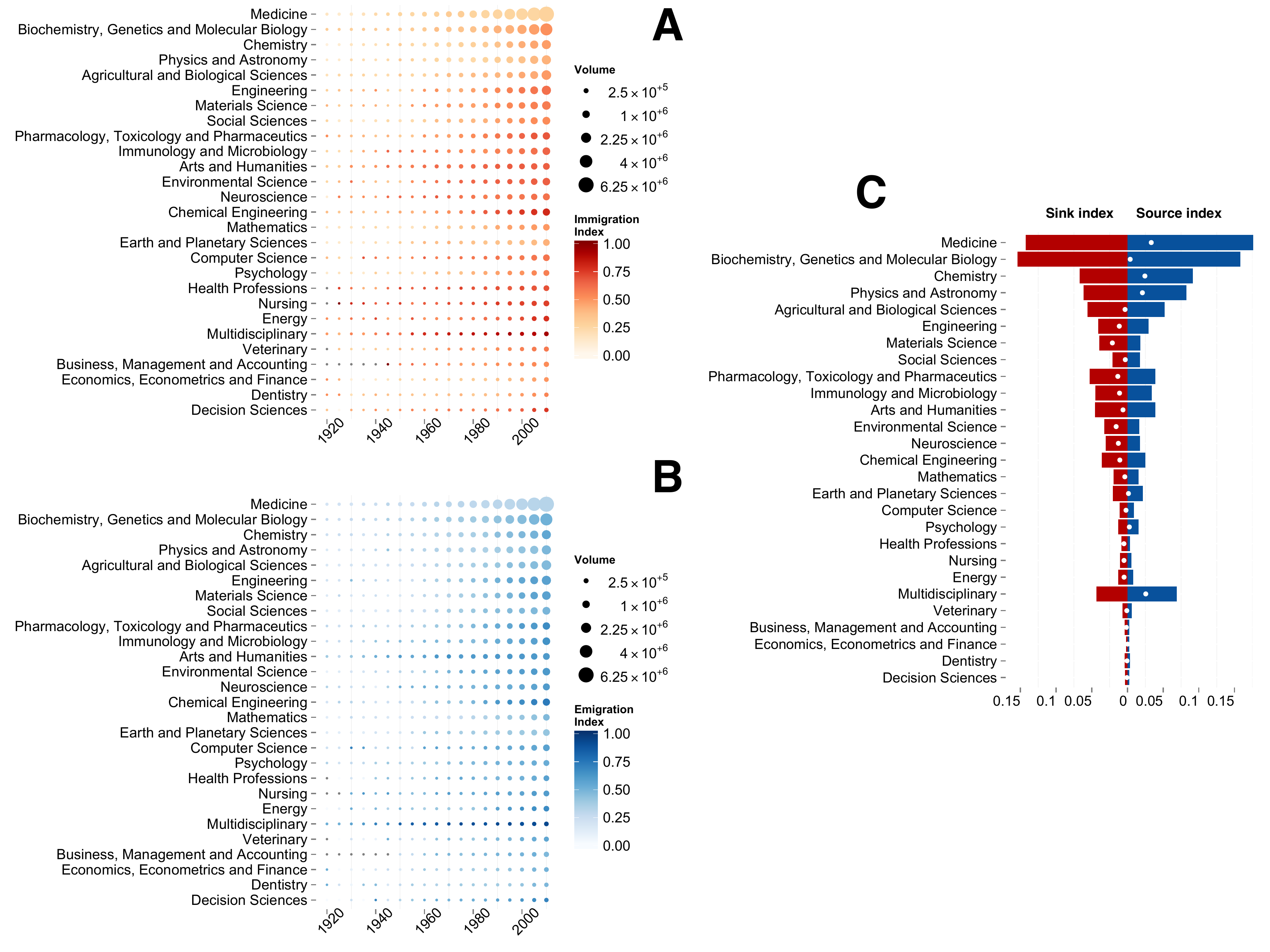}
\caption{\textbf{Incoming and outgoing flows from and to knowledge areas across time.} Immigration (panel A) and emigration (panel B) index (see Eq.~2 and 3) of each area calculated for each temporal snapshot. Here, 0 indicates that the incoming (outgoing) flow of immigrating (emigrating) authors is negligible with respect to the existing authors population in the area, and 1 that the existing population is negligible with respect to the incoming (outgoing) flow of immigrating (emigrating) authors. Size of circles are proportional to the volume of authors in each area, and areas are ordered according to their overall volume over time. (C) Median sink (left, red boxes) and source (right, blue boxes) index (see Eq.~4 and 5) calculated for each area. Both range from 0 -- indicating areas where the incoming (outgoing) flow of authors is negligible with respect to the overall incoming (outgoing) flow --  to 1 -- characterizing areas where the incoming (outgoing) flow of authors dominates the overall incoming (outgoing) flow.}
\label{fig:Fig3}
\end{figure*}

In Fig.~3A--B is shown the evolution of the immigration and emigration across years for each area separately. Noticeably, most knowledge areas exhibit an evolution from an initial phase, where the incoming and outgoing flows of authors are negligible with respect to the existing authors population in the area, to the actual phase where these flows gain more and more importance. Nevertheless, some areas like Medicine, Physics and Astronomy, Chemistry or Mathematics are more secluded than others and partially preserve their isolation in both incoming and outgoing flow after one century. Conversely, a few areas like Nursing and Health Professions already exhibited a relevant outgoing flow almost a century ago. Of particular interest are those areas that were isolated a century ago but that, between '60s and '70s, have undergone a transition and started to both attract researchers from and provide researchers for other areas, such as Computer Science and Environmental Science. In the two decades between '50s and '70s many public and governmental research institutions invested on technological and theoretical investigation attracting, among others, mathematicians, physicists, philosophers and engineers. 
During the same years, the raise of Artificial Intelligence, required cross-disciplinary research at the edge of philosophy of mind, electrical engineering, neurophysiology, social intelligence and applied mathematics, to cite a few. In parallel, an inverse flow begun as well when a variety of disciplines started to take advantages of the new tools and methods provided by this area, like for example the emerging field of Digital Humanities.
In the case of Environmental Science, the diaspora coincides with the revolution of the field in the '60s. In fact, the environmental movements born in that period to protest against chemical companies led to the creation of the U.S. Environmental Protection Agency and to the creation of many new environmental laws that required the development of specific environmental protocols of investigation, involving experts from a wide variety of disciplines.
Fig.~3C shows the median over time of the source and sink indices for each area separately, which give instead a global perspective of incoming and outgoing flows. This allows to see that fields like Medicine and Physics, that seem isolated when analyzed locally, actually serve as sinks and sources of the knowledge diaspora. This means that, even though most research in these areas is carried out by authors who are already in the field, their contribution to the overall flow of knowledge is very relevant. In particular, both areas serve mostly as source of the diaspora, supplying other areas with researchers importing new methods and tools.

\begin{figure}[!ht]
\centering
\includegraphics[width=9cm]{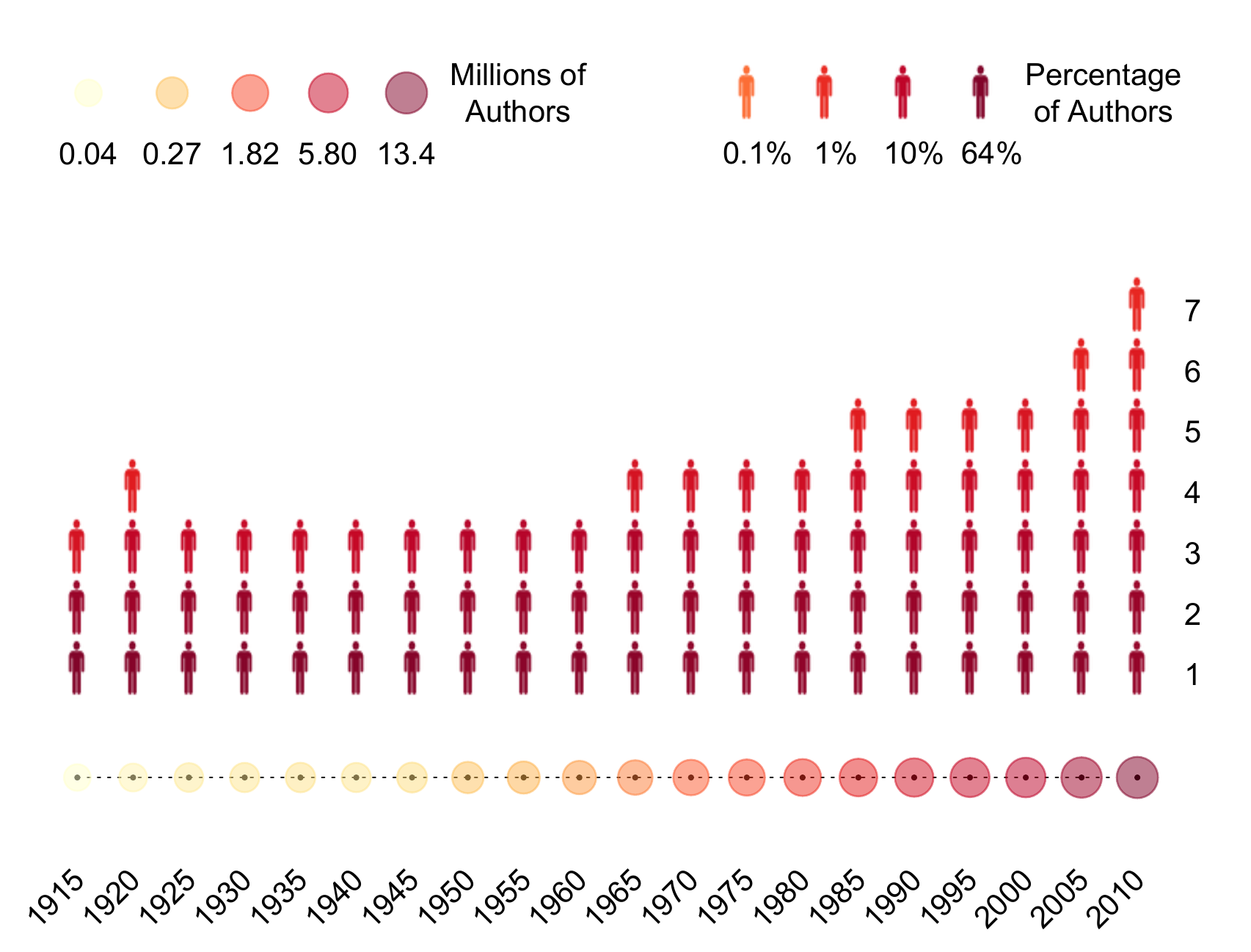}
\caption{\textbf{Authors contribution to different areas.} Each column represents the 99\%-quantile distribution of the number of different scientific areas that an author has published in during the corresponding temporal snapshot. Each icon represents, through its color, the density of authors having published in a given number of areas during a given temporal snapshot. The figure clearly shows that over time authors have increasingly started to publish in more and more scientific areas, i.e. they are becoming more and more multidisciplinary. The circles along the time axis represent the volume of authors that have published during the corresponding temporal snapshot, for reference.}\label{fig:Fig4}
\end{figure}

The knowledge diaspora obliged many researchers to work at the edge of different topics and different areas, driving an increasing trend towards higher trans-disciplinary and multidisciplinary research, in agreement with very recent evidences~\cite{van2015interdisciplinary}. Our data set allows us to quantify also the contribution of authors to different areas during the past 100 years. For each temporal snapshot of the network, we calculate the distribution of the number of different knowledge areas where an author has published in. The evolution of this distribution is shown in Fig.~4 where, as expected, we can observe how authors publish mainly in one area at the beginning of the past century while, over the years, a growing fraction of researchers has begun to produce publications in an increasing number of different areas.

\section{Discussion and conclusions}

We have investigated the evolution of human knowledge across one century by using, as a proxy, the publication patterns of academics in different areas of research. For this purpose, we have used the Microsoft Academic Graph, the largest publicly available data set providing detailed information about academic publications in all areas of knowledge. Our multilayer network map allowed us to model the changes in research interests of academics across time, revealing what we called the ``diaspora of the knowledge''. In fact, we were able to identify disciplines acting as sources or sinks of academics' interest, quantifying their attractiveness across time and revealing fundamental periods in the raise of interest in areas of human knowledge. Noticeably, such periods might be related to crucial historical and political events.
Our results show that, in the last century, a growing number of researchers published papers in an increasing number of disciplines. This clear trend illustrates, in a quantitative way, the perceived growth in the number of authors performing research crossing the boundaries of knowledge areas.

\section*{Author's contributions}
M.D.D and E.O. analyzed the data and performed the analysis. M.D.D., E.O. and A.A. designed the study and wrote the paper. All authors reviewed and approved the complete manuscript.

\section*{Competing interests}
The authors declare no competing interests.

\section*{Acknowledgements}
M.D.D. acknowledges financial support from the Spanish program Juan de la Cierva (IJCI-2014-20225). E.O. was supported by James S.\ McDonnell Foundation.. A.A. acknowledges financial support from ICREA Academia and James S.\ McDonnell Foundation and Spanish MINECO FIS2015-71582.

\appendix

\section{Building the diaspora network}

\begin{figure}[!ht]
\centering
\includegraphics[width=8cm]{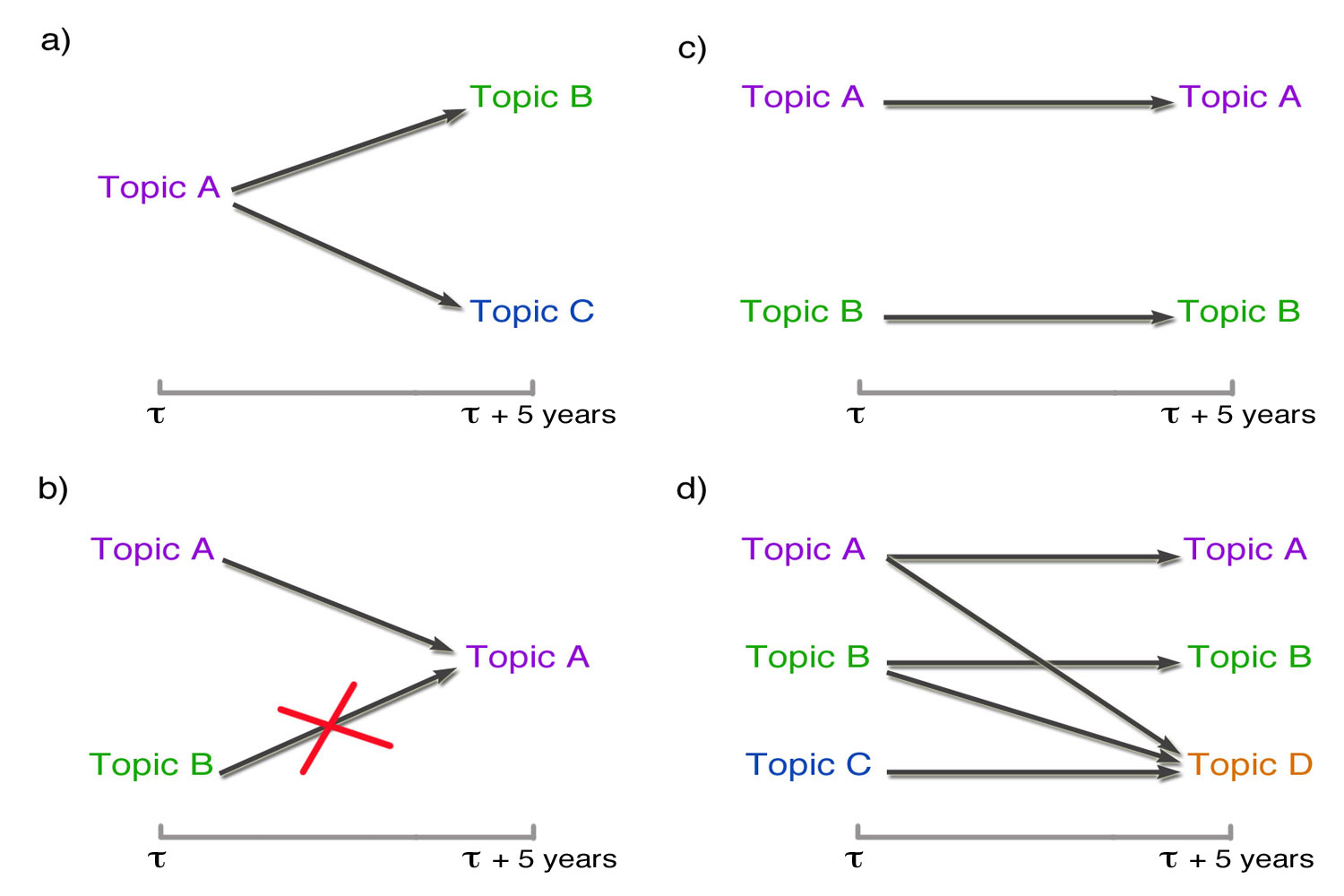}
\caption{\textbf{Knowledge diaspora between areas.} (a) If an author publishes in different topics at time $\tau$ and at time $\tau+\Delta \tau$, we count one transition between all combinations of topics; (b) if an author publishes in topics A and B at time $\tau$, and at time $\tau+\Delta \tau$ again in topic A but not in B anymore, then we consider just one self-transition from topic A to itself; (c) consistently, if an author publishes in topics A and B both at time $\tau$ and at time $\tau+\Delta \tau$, we only count two self-transitions; (d) more generally, if an author publishes in different topics at time $\tau$, and one of them (C) disappears at time $\tau+\Delta \tau$, whereas another (D) appears, since we can not know from with topic at time $\tau$ there is a transition to topic D at time $\tau+\Delta \tau$, we therefore invoke the maximum entropy principle, suggesting that we have to count one transition from any topic at time $\tau$ to topic D.}\label{fig:count}
\end{figure}

Figure~\ref{fig:count} illustrates how we define knowledge diaspora in terms of authors' movements across their research interests.

\section{Categorical edge-bundling visualization of networks}\label{sup:viz}

Visualizing in a clear and informative way the intricate web of transitions between different areas is a challenging problem. When the number of interested nodes, in our case topics or areas, and their interconnections is sufficiently small, chord diagrams~\cite{abel2014quantifying} are suitable candidates. However, if the number of interconnections is too large, chord diagrams might lose their high level of readability. We found a good alternative in edge-bundling visualization~\cite{holten2006hierarchical}, although this approach requires hierarchical data and our network does not exhibit any natural hierarchy, that should instead obtained by applying external algorithms and it would be based on assumptions. Instead, what we wanted to exploit is the intrinsic categorization of authors and papers in areas and topics, while having full control on redirecting edges and place nodes according to our needing. Inspired by Circos visualization~\cite{krzywinski2009circos}, we adopted a circular layout, i.e. embedding on a circle, where categories, in our case the areas of knowledge, are drawn as sectors with different colors. The position of sectors is chosen according to heuristics depending, among other factors, on the modular structure~\cite{newman2012communities} of the network of layers. Nodes, in our case the topics, are placed on a circular layout, close to the sector encoding the area they belong to. Within each sector, nodes are ordered by the logarithm of their strength, to facilitate the identification of important topics and improve the visualization of connections. The size of nodes is rescaled to avoid nodes with radius below or above certain thresholds. The name of topics, i.e. node's label, is shown radially along the direction connecting the node to the center of the circle and both nodes and labels are colored according to the area they belong to, to facilitate readability. Edges are divided into three categories: ``intra-area'' (encoding connections among topics within the same area regardless of direction), ``cross-area out-going'' (encoding connections going from a topic to other topics in a different area) and ``cross-area in-going'' (encoding connections going to a topic from other topics in a different area). Intra-area edges are spline curves placed in the space between the sectors and the nodes, colored by using the color of the underlying area, allowing to gain insight about the diaspora within the same area. Cross-area edges are spline curves calculated by using five points, in addition to the positions of origin and destination nodes: 1) in front of the origin node, belonging to a ``zero-level'' circle with smaller radius than that where nodes are placed; 2) on a ``first-level'' circle with a smaller radius than the previous one, through a point whose position is on the right of the barycenter of the underlying sector and slightly displaced towards the sector; 3) on a ``second-level'' circle, through a point that is aligned with the barycenter of the two endpoints; 4) again on the first-level circle, through a point whose position is on the left of the barycenter of the underlying sector and slightly displaced towards the center of the circle, at variance with the third point; 5) one, on the zero-level circle, in front of the destination node. The displacement of a small angle to the right and to the left allows to separate the out-going and the in-going edges, respectively, as well as the small displacement along the radial direction facilitates the distinguishability of flow directionality, with out-going flow collected into a point closer to the sector and in-going flow collected into a point closer to the center of the circle. The color of each edge is calculated by interpolating the colors of the endpoints, while giving more weight to the color of the destination. The width of the edges is proportional to their weight, i.e. in our case the volume of authors between the endpoint topics and their transparency is regulated by the Euclidean distance between the connected nodes according to their position in the circular layout.

\clearpage
\bibliographystyle{apsrev} 
\bibliography{knowledge_diaspora}

\begin{thebibliography}{23}
\expandafter\ifx\csname natexlab\endcsname\relax\def\natexlab#1{#1}\fi
\expandafter\ifx\csname bibnamefont\endcsname\relax
  \def\bibnamefont#1{#1}\fi
\expandafter\ifx\csname bibfnamefont\endcsname\relax
  \def\bibfnamefont#1{#1}\fi
\expandafter\ifx\csname citenamefont\endcsname\relax
  \def\citenamefont#1{#1}\fi
\expandafter\ifx\csname url\endcsname\relax
  \def\url#1{\texttt{#1}}\fi
\expandafter\ifx\csname urlprefix\endcsname\relax\def\urlprefix{URL }\fi
\providecommand{\bibinfo}[2]{#2}
\providecommand{\eprint}[2][]{\url{#2}}

\bibitem[{\citenamefont{Boyack and B{\"o}rner}(2003)}]{boyack2003indicator}
\bibinfo{author}{\bibfnamefont{K.~W.} \bibnamefont{Boyack}} \bibnamefont{and}
  \bibinfo{author}{\bibfnamefont{K.}~\bibnamefont{B{\"o}rner}},
  \bibinfo{journal}{Journal of the American Society for Information Science and
  Technology} \textbf{\bibinfo{volume}{54}}, \bibinfo{pages}{447}
  (\bibinfo{year}{2003}).

\bibitem[{\citenamefont{Ma et~al.}(2015)\citenamefont{Ma, Mondrag{\'o}n, and
  Latora}}]{ma2015anatomy}
\bibinfo{author}{\bibfnamefont{A.}~\bibnamefont{Ma}},
  \bibinfo{author}{\bibfnamefont{R.~J.} \bibnamefont{Mondrag{\'o}n}},
  \bibnamefont{and} \bibinfo{author}{\bibfnamefont{V.}~\bibnamefont{Latora}},
  \bibinfo{journal}{Proceedings of the National Academy of Sciences} p.
  \bibinfo{pages}{{In Press}} (\bibinfo{year}{2015}).

\bibitem[{\citenamefont{Etzkowitz and Leydesdorff}(1995)}]{etzkowitz1995triple}
\bibinfo{author}{\bibfnamefont{H.}~\bibnamefont{Etzkowitz}} \bibnamefont{and}
  \bibinfo{author}{\bibfnamefont{L.}~\bibnamefont{Leydesdorff}},
  \bibinfo{journal}{EASST Review} \textbf{\bibinfo{volume}{14}},
  \bibinfo{pages}{14} (\bibinfo{year}{1995}).

\bibitem[{\citenamefont{Leydesdorff and
  Etzkowitz}(1998)}]{leydesdorff1998triple}
\bibinfo{author}{\bibfnamefont{L.}~\bibnamefont{Leydesdorff}} \bibnamefont{and}
  \bibinfo{author}{\bibfnamefont{H.}~\bibnamefont{Etzkowitz}},
  \bibinfo{journal}{Science and public policy} \textbf{\bibinfo{volume}{25}},
  \bibinfo{pages}{195} (\bibinfo{year}{1998}).

\bibitem[{\citenamefont{Etzkowitz and
  Leydesdorff}(2000)}]{etzkowitz2000dynamics}
\bibinfo{author}{\bibfnamefont{H.}~\bibnamefont{Etzkowitz}} \bibnamefont{and}
  \bibinfo{author}{\bibfnamefont{L.}~\bibnamefont{Leydesdorff}},
  \bibinfo{journal}{Research policy} \textbf{\bibinfo{volume}{29}},
  \bibinfo{pages}{109} (\bibinfo{year}{2000}).

\bibitem[{\citenamefont{Boyack et~al.}(2005)\citenamefont{Boyack, Klavans, and
  B{\"o}rner}}]{boyack2005mapping}
\bibinfo{author}{\bibfnamefont{K.~W.} \bibnamefont{Boyack}},
  \bibinfo{author}{\bibfnamefont{R.}~\bibnamefont{Klavans}}, \bibnamefont{and}
  \bibinfo{author}{\bibfnamefont{K.}~\bibnamefont{B{\"o}rner}},
  \bibinfo{journal}{Scientometrics} \textbf{\bibinfo{volume}{64}},
  \bibinfo{pages}{351} (\bibinfo{year}{2005}).

\bibitem[{\citenamefont{Leydesdorff and Rafols}(2009)}]{leydesdorff2009global}
\bibinfo{author}{\bibfnamefont{L.}~\bibnamefont{Leydesdorff}} \bibnamefont{and}
  \bibinfo{author}{\bibfnamefont{I.}~\bibnamefont{Rafols}},
  \bibinfo{journal}{Journal of the American Society for Information Science and
  Technology} \textbf{\bibinfo{volume}{60}}, \bibinfo{pages}{348}
  (\bibinfo{year}{2009}).

\bibitem[{\citenamefont{Deville et~al.}(2014)\citenamefont{Deville, Wang,
  Sinatra, Song, Blondel, and Barab{\'a}si}}]{deville2014career}
\bibinfo{author}{\bibfnamefont{P.}~\bibnamefont{Deville}},
  \bibinfo{author}{\bibfnamefont{D.}~\bibnamefont{Wang}},
  \bibinfo{author}{\bibfnamefont{R.}~\bibnamefont{Sinatra}},
  \bibinfo{author}{\bibfnamefont{C.}~\bibnamefont{Song}},
  \bibinfo{author}{\bibfnamefont{V.~D.} \bibnamefont{Blondel}},
  \bibnamefont{and} \bibinfo{author}{\bibfnamefont{A.-L.}
  \bibnamefont{Barab{\'a}si}}, \bibinfo{journal}{Scientific reports}
  \textbf{\bibinfo{volume}{4}}, \bibinfo{pages}{4770} (\bibinfo{year}{2014}).

\bibitem[{\citenamefont{Shiffrin and B{\"o}rner}(2004)}]{shiffrin2004mapping}
\bibinfo{author}{\bibfnamefont{R.~M.} \bibnamefont{Shiffrin}} \bibnamefont{and}
  \bibinfo{author}{\bibfnamefont{K.}~\bibnamefont{B{\"o}rner}},
  \bibinfo{journal}{Proceedings of the National Academy of Sciences}
  \textbf{\bibinfo{volume}{101}}, \bibinfo{pages}{5183} (\bibinfo{year}{2004}).

\bibitem[{\citenamefont{B{\"o}rner et~al.}(2004)\citenamefont{B{\"o}rner, Maru,
  and Goldstone}}]{borner2004simultaneous}
\bibinfo{author}{\bibfnamefont{K.}~\bibnamefont{B{\"o}rner}},
  \bibinfo{author}{\bibfnamefont{J.~T.} \bibnamefont{Maru}}, \bibnamefont{and}
  \bibinfo{author}{\bibfnamefont{R.~L.} \bibnamefont{Goldstone}},
  \bibinfo{journal}{Proceedings of the National Academy of Sciences}
  \textbf{\bibinfo{volume}{101}}, \bibinfo{pages}{5266} (\bibinfo{year}{2004}).

\bibitem[{\citenamefont{Ke et~al.}(2015)\citenamefont{Ke, Ferrara, Radicchi,
  and Flammini}}]{ke2015defining}
\bibinfo{author}{\bibfnamefont{Q.}~\bibnamefont{Ke}},
  \bibinfo{author}{\bibfnamefont{E.}~\bibnamefont{Ferrara}},
  \bibinfo{author}{\bibfnamefont{F.}~\bibnamefont{Radicchi}}, \bibnamefont{and}
  \bibinfo{author}{\bibfnamefont{A.}~\bibnamefont{Flammini}},
  \bibinfo{journal}{Proceedings of the National Academy of Sciences} p.
  \bibinfo{pages}{201424329} (\bibinfo{year}{2015}).

\bibitem[{\citenamefont{Sinatra et~al.}(2015)\citenamefont{Sinatra, Deville,
  Szell, Wang, and Barab{\'a}si}}]{sinatra2015century}
\bibinfo{author}{\bibfnamefont{R.}~\bibnamefont{Sinatra}},
  \bibinfo{author}{\bibfnamefont{P.}~\bibnamefont{Deville}},
  \bibinfo{author}{\bibfnamefont{M.}~\bibnamefont{Szell}},
  \bibinfo{author}{\bibfnamefont{D.}~\bibnamefont{Wang}}, \bibnamefont{and}
  \bibinfo{author}{\bibfnamefont{A.-L.} \bibnamefont{Barab{\'a}si}},
  \bibinfo{journal}{Nature Physics} \textbf{\bibinfo{volume}{11}},
  \bibinfo{pages}{791} (\bibinfo{year}{2015}).

\bibitem[{\citenamefont{Holme and Saram{\"a}ki}(2012)}]{holme2012temporal}
\bibinfo{author}{\bibfnamefont{P.}~\bibnamefont{Holme}} \bibnamefont{and}
  \bibinfo{author}{\bibfnamefont{J.}~\bibnamefont{Saram{\"a}ki}},
  \bibinfo{journal}{Phys. Rep.} \textbf{\bibinfo{volume}{519}},
  \bibinfo{pages}{97} (\bibinfo{year}{2012}).

\bibitem[{\citenamefont{Kivel{\"a} et~al.}(2014)\citenamefont{Kivel{\"a},
  Arenas, Barthelemy, Gleeson, Moreno, and Porter}}]{kivela2013multilayer}
\bibinfo{author}{\bibfnamefont{M.}~\bibnamefont{Kivel{\"a}}},
  \bibinfo{author}{\bibfnamefont{A.}~\bibnamefont{Arenas}},
  \bibinfo{author}{\bibfnamefont{M.}~\bibnamefont{Barthelemy}},
  \bibinfo{author}{\bibfnamefont{J.~P.} \bibnamefont{Gleeson}},
  \bibinfo{author}{\bibfnamefont{Y.}~\bibnamefont{Moreno}}, \bibnamefont{and}
  \bibinfo{author}{\bibfnamefont{M.~A.} \bibnamefont{Porter}},
  \bibinfo{journal}{Journal of Complex Networks} \textbf{\bibinfo{volume}{2}},
  \bibinfo{pages}{203} (\bibinfo{year}{2014}).

\bibitem[{\citenamefont{De~Domenico et~al.}(2013)\citenamefont{De~Domenico,
  Sol{\`e}-Ribalta, Cozzo, Kivel{\"a}, Moreno, Porter, G{\`o}mez, and
  Arenas}}]{dedomenico2013mathematical}
\bibinfo{author}{\bibfnamefont{M.}~\bibnamefont{De~Domenico}},
  \bibinfo{author}{\bibfnamefont{A.}~\bibnamefont{Sol{\`e}-Ribalta}},
  \bibinfo{author}{\bibfnamefont{E.}~\bibnamefont{Cozzo}},
  \bibinfo{author}{\bibfnamefont{M.}~\bibnamefont{Kivel{\"a}}},
  \bibinfo{author}{\bibfnamefont{Y.}~\bibnamefont{Moreno}},
  \bibinfo{author}{\bibfnamefont{M.~A.} \bibnamefont{Porter}},
  \bibinfo{author}{\bibfnamefont{S.}~\bibnamefont{G{\`o}mez}},
  \bibnamefont{and} \bibinfo{author}{\bibfnamefont{A.}~\bibnamefont{Arenas}},
  \bibinfo{journal}{Phys. Rev. X} \textbf{\bibinfo{volume}{3}},
  \bibinfo{pages}{041022} (\bibinfo{year}{2013}).

\bibitem[{\citenamefont{Boccaletti et~al.}(2014)\citenamefont{Boccaletti,
  Bianconi, Criado, Del~Genio, G{\'o}mez-Garde{\~n}es, Romance,
  Sendi{\~n}a-Nadal, Wang, and Zanin}}]{boccaletti2014structure}
\bibinfo{author}{\bibfnamefont{S.}~\bibnamefont{Boccaletti}},
  \bibinfo{author}{\bibfnamefont{G.}~\bibnamefont{Bianconi}},
  \bibinfo{author}{\bibfnamefont{R.}~\bibnamefont{Criado}},
  \bibinfo{author}{\bibfnamefont{C.}~\bibnamefont{Del~Genio}},
  \bibinfo{author}{\bibfnamefont{J.}~\bibnamefont{G{\'o}mez-Garde{\~n}es}},
  \bibinfo{author}{\bibfnamefont{M.}~\bibnamefont{Romance}},
  \bibinfo{author}{\bibfnamefont{I.}~\bibnamefont{Sendi{\~n}a-Nadal}},
  \bibinfo{author}{\bibfnamefont{Z.}~\bibnamefont{Wang}}, \bibnamefont{and}
  \bibinfo{author}{\bibfnamefont{M.}~\bibnamefont{Zanin}},
  \bibinfo{journal}{Physics Reports} \textbf{\bibinfo{volume}{544}},
  \bibinfo{pages}{1} (\bibinfo{year}{2014}).

\bibitem[{\citenamefont{Sinha et~al.}(2015)\citenamefont{Sinha, Shen, Song, Ma,
  Eide, Hsu, and Wang}}]{sinha2015overview}
\bibinfo{author}{\bibfnamefont{A.}~\bibnamefont{Sinha}},
  \bibinfo{author}{\bibfnamefont{Z.}~\bibnamefont{Shen}},
  \bibinfo{author}{\bibfnamefont{Y.}~\bibnamefont{Song}},
  \bibinfo{author}{\bibfnamefont{H.}~\bibnamefont{Ma}},
  \bibinfo{author}{\bibfnamefont{D.}~\bibnamefont{Eide}},
  \bibinfo{author}{\bibfnamefont{B.-j.~P.} \bibnamefont{Hsu}},
  \bibnamefont{and} \bibinfo{author}{\bibfnamefont{K.}~\bibnamefont{Wang}}, in
  \emph{\bibinfo{booktitle}{Proceedings of the 24th International Conference on
  World Wide Web Companion}} (\bibinfo{organization}{International World Wide
  Web Conferences Steering Committee}, \bibinfo{year}{2015}), pp.
  \bibinfo{pages}{243--246}.

\bibitem[{\citenamefont{Martin}(2015)}]{martin2015s}
\bibinfo{author}{\bibfnamefont{J.~D.} \bibnamefont{Martin}},
  \bibinfo{journal}{Physics in Perspective} \textbf{\bibinfo{volume}{17}},
  \bibinfo{pages}{3} (\bibinfo{year}{2015}).

\bibitem[{\citenamefont{Van~Noorden}(2015)}]{van2015interdisciplinary}
\bibinfo{author}{\bibfnamefont{R.}~\bibnamefont{Van~Noorden}},
  \bibinfo{journal}{Nature} \textbf{\bibinfo{volume}{525}},
  \bibinfo{pages}{306} (\bibinfo{year}{2015}).

\bibitem[{\citenamefont{Abel and Sander}(2014)}]{abel2014quantifying}
\bibinfo{author}{\bibfnamefont{G.~J.} \bibnamefont{Abel}} \bibnamefont{and}
  \bibinfo{author}{\bibfnamefont{N.}~\bibnamefont{Sander}},
  \bibinfo{journal}{Science} \textbf{\bibinfo{volume}{343}},
  \bibinfo{pages}{1520} (\bibinfo{year}{2014}).

\bibitem[{\citenamefont{Holten}(2006)}]{holten2006hierarchical}
\bibinfo{author}{\bibfnamefont{D.}~\bibnamefont{Holten}},
  \bibinfo{journal}{Visualization and Computer Graphics, IEEE Transactions on}
  \textbf{\bibinfo{volume}{12}}, \bibinfo{pages}{741} (\bibinfo{year}{2006}).

\bibitem[{\citenamefont{Krzywinski et~al.}(2009)\citenamefont{Krzywinski,
  Schein, Birol, Connors, Gascoyne, Horsman, Jones, and
  Marra}}]{krzywinski2009circos}
\bibinfo{author}{\bibfnamefont{M.}~\bibnamefont{Krzywinski}},
  \bibinfo{author}{\bibfnamefont{J.}~\bibnamefont{Schein}},
  \bibinfo{author}{\bibfnamefont{I.}~\bibnamefont{Birol}},
  \bibinfo{author}{\bibfnamefont{J.}~\bibnamefont{Connors}},
  \bibinfo{author}{\bibfnamefont{R.}~\bibnamefont{Gascoyne}},
  \bibinfo{author}{\bibfnamefont{D.}~\bibnamefont{Horsman}},
  \bibinfo{author}{\bibfnamefont{S.~J.} \bibnamefont{Jones}}, \bibnamefont{and}
  \bibinfo{author}{\bibfnamefont{M.~A.} \bibnamefont{Marra}},
  \bibinfo{journal}{Genome research} \textbf{\bibinfo{volume}{19}},
  \bibinfo{pages}{1639} (\bibinfo{year}{2009}).

\bibitem[{\citenamefont{Newman}(2012)}]{newman2012communities}
\bibinfo{author}{\bibfnamefont{M.~E.} \bibnamefont{Newman}},
  \bibinfo{journal}{Nature Physics} \textbf{\bibinfo{volume}{8}},
  \bibinfo{pages}{25} (\bibinfo{year}{2012}).

\end{thebibliography}

\end{document}